\documentclass[prl,aps,showpacs,twocolumn,superscriptaddress]{revtex4}
\usepackage{graphicx}
\usepackage{amssymb}
\usepackage{amsmath}
\usepackage{color}

\newcommand{\beq}{\begin{eqnarray}}
\newcommand{\eeq}{\end{eqnarray}}

\usepackage{psfrag}

\begin{document}

\title{Thermodynamics of $SU(3)$ Gauge Theory from Gradient Flow}

\author{Masayuki~Asakawa}
\email{yuki@phys.sci.osaka-u.ac.jp}
\affiliation{Department of Physics, Osaka University, Toyonaka, Osaka 560-0043,
Japan}

\author{Tetsuo~Hatsuda}
\email{thatsuda@riken.jp}
\affiliation{Theoretical Research Division, Nishina Center, RIKEN, Wako
351-0198, Japan}
\affiliation{Kavli IPMU (WPI), The University of Tokyo, Chiba 606-8502, Japan}

\author{Etsuko~Itou}
\email{eitou@post.kek.jp}
\affiliation{High Energy Accelerator Research Organisation (KEK), Tsukuba
305-0801, Japan}

\author{Masakiyo~Kitazawa}
\email{kitazawa@phys.sci.osaka-u.ac.jp}
\affiliation{Department of Physics, Osaka University, Toyonaka, Osaka 560-0043,
Japan}

\author{Hiroshi~Suzuki}
\email{hsuzuki@phys.kyushu-u.ac.jp}
\affiliation{Department of Physics, Kyushu University, 6-10-1 Hakozaki,
Higashi-ku, Fukuoka, 812-8581, Japan}

\collaboration{FlowQCD Collaboration}

\begin{abstract}
A novel method to study the bulk thermodynamics in lattice gauge theory is
proposed on the basis of the Yang--Mills gradient flow with a fictitious
time~$t$. The energy density~$\varepsilon$ and the pressure~$P$ of $SU(3)$
gauge theory at fixed temperature are calculated directly on
$32^3\times(6,8,10)$ lattices from the thermal average of the well-defined
energy-momentum tensor $T_{\mu\nu}^R(x)$ obtained by the gradient flow. It is
demonstrated that the continuum limit can be taken in a controlled manner from
the $t$-dependence of the flowed data.
\end{abstract}

\maketitle


The symmetric energy momentum tensor (EMT), $T_{\mu\nu}$, which is the generator
of the Poincar\'e transformations, is a fundamental operator in quantum field
theory~\cite{Caracciolo:1990emt}.
Since $T_{00}$, $T_{i0}$, and~$T_{ij}$
correspond to the energy density, the momentum density, and the momentum-flux
density, respectively, 
the EMT and its correlation functions provide useful
information on the bulk and transport properties at finite temperature~($T$).
For example, the energy density~$\varepsilon$ and the pressure~$P$ are given
by~$\langle T_{00}\rangle$ and~$\langle T_{11,22,33}\rangle$, respectively,
with~$\langle\cdot\rangle$ being the thermal average. Also, the shear
viscosity~$\eta$ can be extracted from the two-point correlation,
$\langle T_{12}(x)T_{12}(y)\rangle$. In quantum chromodynamics~(QCD), these
observables are particularly important in formulating the relativistic
hydrodynamics for the quark-gluon plasma~\cite{Romatschke:2009im}. Therefore,
high precision and non-perturbative evaluation of the $n$-point EMT
correlations in lattice QCD is called for.

To calculate such correlations in numerical lattice simulations, we first need
to define proper EMT on the lattice which is ultra-violet (UV) finite and is
conserved in the continuum limit. Such a construction is not a trivial task due
to the explicit breaking of the Poincar\'e invariance on the lattice. (See
Refs.~\cite{Giusti:2012yj,Robaina:2013zmb,Giusti:2013sqa,Giusti:2013mxa} for
recent developments.) This is the reason why $\varepsilon$ and~$P$ at
finite~$T$ have been mainly studied by an indirect ``integral method'' without
the explicit use of the EMT~\cite{Lombardo:2012ix}.

Recently, one of the present authors has shown that the proper EMT keeping
all the nice features can be naturally constructed~\cite{Suzuki:2013gza} on the
basis of the Yang--Mills gradient
flow~\cite{Luscher:2010iy,Luscher:2011bx,Luscher:2013vga}. (See also related
works, Refs.~\cite{Borsanyi:2012zs,Borsanyi:2012zr,Fodor:2012td,%
Fritzsch:2013je,Luscher:2013cpa}.) 
In this Letter, we demonstrate,  for the first time, 
that the thermal $SU(3)$ gauge theory can be studied 
by the direct lattice measurement of the proper EMT by
 considering $\varepsilon$ and $P$ as examples.
The key idea is to represent the EMT in the
continuum limit by UV-finite and local operators obtained from the
gradient flow.
 Then, by taking the limit of small flow time and small lattice
spacing in an appropriate way, as discussed later, accurate
thermodynamic observables are obtained with modest statistics.

Let us first recapitulate the basic idea of~Ref.~\cite{Suzuki:2013gza} in the
continuum space-time. The Yang--Mills gradient flow is a deformation of the
gauge configuration $A_\mu(x)$ along a fictitious Euclidean time~$t$;
$\partial_tB_\mu(t,x)=D_\nu G_{\nu\mu}(t,x)$ with $B_\mu(t=0,x)=A_\mu(x)$,
where $D_\mu$ and $G_{\mu\nu}(t,x)$ are the covariant derivative and the field
strength of the flowed gauge field~$B_\mu(t,x)$, respectively. The color
indices are suppressed for simplicity. A salient feature of the gradient flow
is its UV finiteness: Any correlation functions of $B_{\mu_1}(t_1,x_1)$,
$B_{\mu_2}(t_2,x_2)$, \dots\ for~$t_i>0$ are UV finite without the wave function
renormalization if they are written in terms of the renormalized
coupling~\cite{Luscher:2011bx}. This is owing to the fact that the diffusion
in~$t$ naturally introduces a proper-time regulator of the form $e^{-tp^2}$,
where $p$~denotes a typical loop momentum. In particular, the correlation
functions are free from UV divergences even at the equal-point,
$(t_1,x_1)=(t_2, x_2)=\dotsb$ for positive~$t_i$. For example, the following
gauge-invariant local products of dimension~$4$ are UV finite for~$t>0$:
$U_{\mu\nu}(t,x)\equiv G_{\mu\rho}(t,x)G_{\nu\rho}(t,x)
-\frac{1}{4}\delta_{\mu\nu}G_{\rho\sigma}(t,x)G_{\rho\sigma}(t,x)$
and~$E(t,x)\equiv\frac{1}{4}G_{\mu\nu}(t,x)G_{\mu\nu}(t,x)$.

For $t\to0_+$, local products of flowed fields can be expanded in terms of
four-dimensional renormalized local operators with increasing
dimensions~\cite{Luscher:2011bx}: The expansion coefficients are governed by
the renormalization group equation and their small $t$ behavior can be
calculated by perturbation theory thanks to the asymptotic freedom. For the
operators mentioned above, we have~\cite{Suzuki:2013gza,DelDebbio:2013zaa}
\begin{align}
   U_{\mu\nu}(t,x)
   &=\alpha_U(t)\left[
   T_{\mu\nu}^R(x)-\frac{1}{4}\delta_{\mu\nu}T_{\rho\rho}^R(x)\right]
   +O(t),
\label{eq:(2)}\\
   E(t,x)
   &=\left\langle E(t,x)\right\rangle_0
   +\alpha_E(t)T_{\rho\rho}^R(x)
   +O(t),
\label{eq:(3)}
\end{align}
where $\langle\cdot\rangle_0$ is vacuum expectation value and $T_{\mu\nu}^R(x)$
is the correctly-normalized conserved EMT with its vacuum expectation value
subtracted. Abbreviated are the contributions from the operators of
dimension~$6$ or higher, which are suppressed for small~$t$.

Combining relations Eqs.~(\ref{eq:(2)}) and~(\ref{eq:(3)}), we have
\begin{align}
   &T_{\mu\nu}^R(x)
\notag\\
   &=\lim_{t\to0}\left\{\frac{1}{\alpha_U(t)}U_{\mu\nu}(t,x)
   +\frac{\delta_{\mu\nu}}{4\alpha_E(t)}
   \left[E(t,x)-\left\langle E(t,x)\right\rangle_0 \right]\right\},
\label{eq:(4)}
\end{align}
where the perturbative coefficients are found to be~\cite{Suzuki:2013gza}
\begin{align}
   \alpha_U(t)
   &=\Bar{g}(1/\sqrt{8t})^2
   \left[1+2b_0\Bar{s}_1\Bar{g}(1/\sqrt{8t})^2+O(\Bar{g}^4)\right],
\label{eq:(5)}
\\
   \alpha_E(t)
   &=\frac{1}{2b_0}\left[1+2b_0\Bar{s}_2
   \Bar{g}(1/\sqrt{8t})^2+O(\Bar{g}^4)\right].
\label{eq:(6)}
\end{align}
Here $\Bar{g}(q)$ denotes the running gauge coupling in the
$\overline{\text{MS}}$ scheme with the choice, $q=1/\sqrt{8t}$, and
$\Bar{s}_1=\frac{7}{22}+\frac{1}{2}\gamma_E-\ln2\simeq -0.08635752993$,
$\bar{s}_2=\frac{21}{44}-\frac{b_1}{2b_0^2}=\frac{27}{484}\simeq0.05578512397$,
with 
$b_0=\frac{1}{(4\pi)^2}\frac{11}{3}N_c$,
$b_1=\frac{1}{(4\pi)^4}\frac{34}{3}N_c^2$,
and~$N_c=3$. Note that a non-perturbative determination of~$\alpha_{U,E}(t)$ is
also proposed recently~\cite{DelDebbio:2013zaa}.

The formula Eq.~(\ref{eq:(4)}) indicates that $T_{\mu\nu}^R(x)$ can be obtained
by the small $t$ limit of the gauge-invariant local operators defined
through the gradient flow. There are two important observations:
(i)~The right-hand side of~Eq.~(\ref{eq:(4)}) is independent of the
regularization because of its UV finiteness, so that one can take, e.g.\ the
lattice regularization scheme;
(ii)~since flowed fields at~$t>0$ depend on the fundamental fields at~$t=0$ in
the space-time region of radius~$\simeq\sqrt{8t}$, the statistical noise in
calculating the right hand side of~Eq.~(\ref{eq:(4)}) is suppressed for
finite~$t$.

Our procedure to calculate the EMT on the lattice has the
following four steps:\\
\textbf{Step 1:} Generate gauge configurations at~$t=0$ on a space-time lattice
with the lattice spacing~$a$ and the lattice size~$N_s^3\times N_\tau$.\\
\textbf{Step 2:} Solve the gradient flow for each configuration to obtain the
flowed link variables in the fiducial window, $a\ll\sqrt{8t}\ll R$. Here,
$R$~is an infrared cutoff scale such as~$\Lambda_{\text{QCD}}^{-1}$
or~$T^{-1}=N_\tau a$. The first (second) inequality is necessary to suppress 
finite $a$ corrections (non-perturbative corrections and finite volume
corrections).\\
\textbf{Step 3:} Construct $U_{\mu\nu}(t,x)$ and~$E(t,x)$
in~Eqs.~(\ref{eq:(2)}) and~(\ref{eq:(3)}) in terms of the flowed link variables
and average over the gauge configurations at each~$t$.\\
\textbf{Step 4:} 
Carry out an extrapolation toward $(a,t)=(0,0)$, first $a\to0$ and then $t\to0$ under the condition in \textbf{Step 2}.

The thermodynamic quantities
are obtained from the diagonal elements of the EMT:  
A combination of $\varepsilon$ and~$P$ 
called the interaction measure~$\Delta$
 is related to the trace of the EMT (the trace
anomaly):
\begin{equation}
   \Delta=\varepsilon-3P
   =-\left\langle T_{\mu\mu}^R(x)\right\rangle.
\label{eq:(7)}
\end{equation}
Also, the entropy density~$s$ at zero chemical potential reads
\begin{equation}
   sT=\varepsilon+P
   =-\langle T_{00}^R(x)\rangle
   +\frac{1}{3}\sum_{i=1,2,3}\langle T_{ii}^R(x)\rangle.
\label{eq:(8)}
\end{equation}

To demonstrate that the above four Steps can be indeed pursued, we consider the
$SU(3)$ gauge theory defined on a four-dimensional Euclidean lattice, whose
thermodynamics has been extensively studied by the integral
method~\cite{Boyd:1996bx,Okamoto:1999hi,Umeda:2008bd,Borsanyi:2012ve}. For
simplicity, we consider the Wilson plaquette gauge action under the periodic
boundary condition on~$N_s^3\times N_\tau=32^3\times(6,8,10)$ lattices with
several different $\beta=6/g_0^2$ ($g_0$~being the bare coupling constant).
Gauge configurations are generated by the pseudo-heatbath algorithm with the
over-relaxation, mixed in the ratio of~$1:5$. We call one pseudo-heatbath
update sweep plus five over-relaxation sweeps as a ``Sweep''. To eliminate the
autocorrelation, we take $200$--$500$ Sweeps between measurements. The number
of gauge configurations for the measurements at finite~$T$ is~$300$.
Statistical errors are estimated by the jackknife method.

\begin{table}[t]
\begin{center}
\begin{tabular}{|c||c|c|c||c|}
\hline
$N_\tau$ & 6 & 8 & 10 & $T/T_c$ \\
\hline
         & 6.20 & 6.40 & 6.56 &  1.65 \\ 
$\beta$  & 6.02 & 6.20 & 6.36 &  1.24 \\
         & 5.89 & 6.06 & 6.20 &  0.99 \\
\hline
\end{tabular}
\caption{Values of $\beta$ and~$N_\tau$ for each temperature.}
\label{table:Nt_beta}
\end{center}
\end{table}

To relate $T/T_c$ and corresponding $\beta$ for each~$N_\tau$, we first use the
relation between $a/r_0$ ($r_0$ is the Sommer scale) and~$\beta$ given by the
ALPHA Collaboration~\cite{Guagnelli:1998ud}. The resultant values of
$Tr_0=[N_\tau(a/r_0)]^{-1}$ are then converted to~$T/T_c$ by using the result
at~$\beta=6.20$ in~Ref.~\cite{Boyd:1996bx}. Nine combinations
of~$(N_\tau,\beta)$ and corresponding $T/T_c$ obtained by this procedure are
shown in~Table~\ref{table:Nt_beta}.

The gradient flow in the $t$-direction is obtained by solving the ordinary
first-order differential equation. We utilize the modified second-order Runge--Kutta method 
in which the error per step ($t \rightarrow t+\epsilon$) is $O(\epsilon^3)$.
We take $\epsilon=0.025$, and confirm that the accumulation errors are sufficiently smaller
than the statistical errors.

To extract the EMT from~Eq.~(\ref{eq:(4)}), we measure
$G^a_{\mu\rho}(t,x)G^a_{\nu\rho}(t,x)$ written in terms of the clover leaf
representation on the lattice. To subtract out the $T=0$ contribution,
$\langle E(t,x)\rangle_0$, we carry out simulations on a $32^4$ lattice for
each~$\beta$ in~Table~\ref{table:Nt_beta}. Note that this vacuum subtraction is
required for
 the trace anomaly~$\Delta$, but
 not for 
the entropy density~$s$.
For~$\Bar{g}$ in $\alpha_U(t)$ and~$\alpha_E(t)$ in~Eqs.~(\ref{eq:(5)})
and~(\ref{eq:(6)}), we use the four-loop running coupling with the scale
parameter determined by the ALPHA Collaboration,
$\Lambda_{\overline{\text{MS}}}=0.602(48)/r_0$~\cite{Capitani:1998mq}. We
confirmed the previous finding~\cite{Luscher:2010iy} that the lattice data of
$t^2\langle E(t,x)\rangle_0$ in the fiducial window matches quite well with its
perturbative estimate in the continuum,
$t^2\langle E(t,x)\rangle_0\simeq3\Bar{g}^2/(4\pi)^2
[1+1.0978\Bar{g}(1/\sqrt{8t})^2/(4\pi)]$ with the four-loop running coupling
and the above $\Lambda_{\overline{\text{MS}}}$.

\begin{figure}[t]
\begin{center}
\includegraphics[scale=0.45]{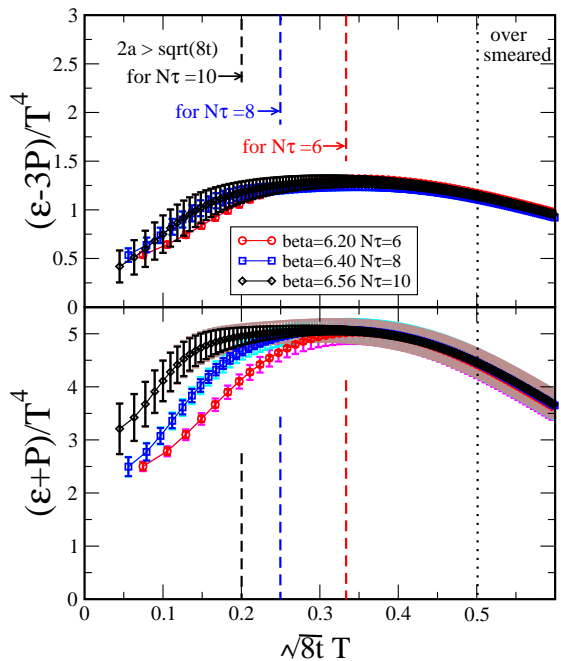}
\caption{Flow time dependence of the dimensionless interaction measure (top
panel) and the dimensionless entropy density (bottom panel) for different
lattice spacings at fixed~$T/T_c=1.65$. The circles (red) the squares (blue),
and the diamonds (black) correspond to~$N_\tau=6$, $8$, and~$10$, respectively.
The bold error bars denote the statistical errors, while the thin error bars
(brown, cyan, and magenta) include both statistical and systematic errors.}
\label{fig:raw-data-1.65Tc}
\end{center}
\end{figure}

Shown in~Fig.~\ref{fig:raw-data-1.65Tc} is our results for the
dimensionless interaction measure ($\Delta/T^4=(\varepsilon-3P)/T^4$) and the
dimensionless entropy density ($s/T^3=(\varepsilon+P)/T^4$) at~$T=1.65T_c$ as a
function of the dimensionless flow parameter~$\sqrt{8t}T$. The bold bars denote
the statistical errors, while the thin (light color) bars show the statistical
and systematic errors including the uncertainty
of~$\Lambda_{\overline{\text{MS}}}$. 
In the small $t$ region, the statistical error is
dominant  for both $\Delta/T^4$ and~$s/T^3$, while in the large $t$ region
the systematic error from~$\Lambda_{\overline{\text{MS}}}$ becomes significant
for~$s/T^3$.
 For instance, the statistical (systematic) errors of the data for $N_\tau=8$ are
 $2.5\%$ ($0.11\%$) for $\Delta/T^4$
and $0.83\%$ ($4.4\%$) for $s/T^3$ at $\sqrt{8t}T=0.40$.

The fiducial window discussed in~\textbf{Step 2} is indicated by the dashed
lines in~Fig.~\ref{fig:raw-data-1.65Tc}. The lower limit, beyond
which the lattice discretization error grows, is set to
be~$\sqrt{8t_{\rm min}}=2a$, 
where we consider the size~$2a$ of our clover
leaf operator. The upper limit, beyond
which the smearing by the gradient flow exceeds the temporal lattice size, is
set to be $\sqrt{8t_{\rm max}}=1/(2T)=N_\tau a/2$.

The data in~Fig.~\ref{fig:raw-data-1.65Tc} show, within the error bars, that
(i)~the plateau appears inside the preset fiducial window
($2/N_\tau<\sqrt{8t}T<1/2$) for each~$N_\tau$, and (ii)~the plateau extends to
the smaller~$t$ region as $N_\tau$ increases or equivalently as $a$ decreases.
Similar plateaus as in~Fig.~\ref{fig:raw-data-1.65Tc}
also appear inside the fiducial window for other temperatures, $T/T_c= 1.24$
and~$0.99$, with comparable error bars. These features imply that 
{the double extrapolation $(a,t)\to(0,0)$ in
$\textbf{Step 4}$  is indeed doable.}

\begin{figure}[t]
\begin{center}
\includegraphics[scale=0.45]{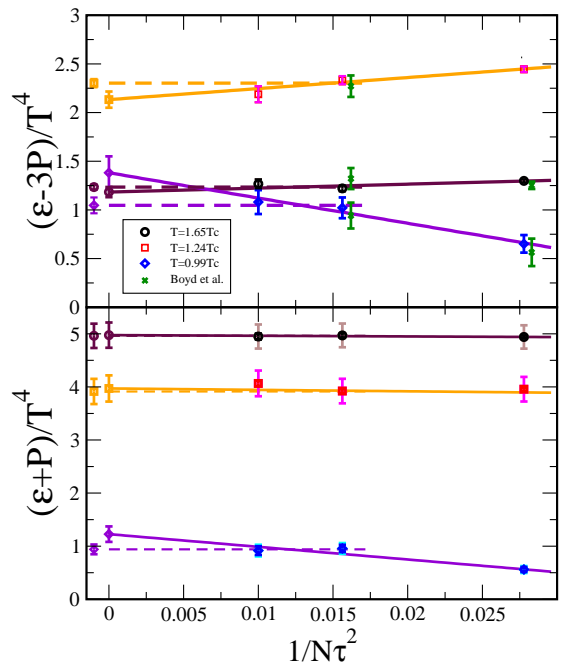}
\caption{Continuum extrapolation of the thermodynamic quantities
for~$T/T_c=1.65$, $1.24$, and~$0.99$. Solid lines and dashed lines correspond
to the three-point linear fit and two-point constant fit as a function
of~$1/N_\tau^2$, respectively. Extrapolated values of the former (latter) are
shown at~$1/N_\tau^2=0$ ($1/N_\tau^2=-0.001$). The cross symbols in the top
panel are the data of~Ref.~\cite{Boyd:1996bx} with the same lattice setup.}
\label{fig:cont-lim}
\end{center}
\end{figure}

Our lattice results at fixed~$T$ with three different lattice spacings allow us
to take the continuum limit. First, we pick up a flow time $\sqrt{8t}T=0.40$
which is in the middle of the fiducial window. Then we extract $\Delta/T^4$
and~$s/T^3$ for each set of~$N_\tau$ and~$\beta$. We have checked that different
choices of~$t$ do not change the final results within the error bar as long as
it is in the plateau region. In~Fig.~\ref{fig:cont-lim}, resultant values
taking into account the statistical errors (bold error bars) and the
statistical plus systematic errors (thin error bars) are shown. The lattice
data for~$\Delta/T^4$ with the same lattice setup at~$N_\tau=6$ and~$8$
in~Ref.~\cite{Boyd:1996bx} are also shown by the cross (green) symbols in the
top panel; 
the statistical error of our result on 
$32^3\times8$ lattice for $\beta=6.4$ ($\beta=6.2$) is about 
$3.33$ ($2.69$) times smaller than the one in Ref.~\cite{Boyd:1996bx} 
obtained on the same lattice. In this way, 
our results with $300$ gauge configurations have substantially
smaller error bars at these points.

The horizontal axis of~Fig.~\ref{fig:cont-lim}, $1/N_\tau^2$, is a variable
suited for making continuum extrapolation of the thermodynamic
quantities~\cite{Boyd:1996bx}. We consider two extrapolation: A linear
fit with the data at~$N_\tau=6$, $8$, and~$10$ (the solid lines
in~Fig.~\ref{fig:cont-lim}), and a constant fit with the data at~$N_\tau=8$
and~$10$ (the dashed lines in~Fig.~\ref{fig:cont-lim}). In both fits, the
correlation between the errors due to the common systematic error
from~$\Lambda_{\overline{\text{MS}}}$ is taken into account. The former fit is
used to determine the central value in the continuum limit whose error is
within~$\pm12\%$ even at our lowest temperature. The latter is used to estimate
the systematic error from the scaling violation whose typical size
is~$\pm5\%$ at high temperature and~$\pm25\%$ at low temperature.

We have analyzed various systematic errors; the perturbative expansion
of~$\alpha_{U,E}(t)$, the running coupling~$\Bar{g}$, the scale parameter, and
the continuum extrapolation. We found that the dominant errors in the present
lattice setup are those from~$\Lambda_{\overline{\text{MS}}}$ and the continuum
extrapolation, which are included in~Fig.~\ref{fig:cont-lim}. To reduce these
systematic errors, finer lattices are quite helpful: They make the plateau
in~$\sqrt{8t}T$ wider by reducing the lower limit of the fiducial window, so
that the continuum extrapolation becomes easier. 
We also note that our continuum extrapolation with fixed 
$N_s=32$ would receive the finite volume effect especially for 
lower $T$ \cite{Borsanyi:2012zs}.
Larger aspect ratio $N_s/N_\tau$ would be helpful 
to guarantee the thermodynamic limit.
Moreover,
the scale setting procedure could be improved to have better accuracy: Instead
of the Sommer scale~$r_0$ adopted in this Letter, more precise scale
determination, e.g.\ by~$t_0$ or~$\omega_0$ in the gradient flow
approach~\cite{Luscher:2010iy,Fodor:2012td}, will be useful.

\begin{figure}[t]
\begin{center}
\includegraphics[scale=0.45]{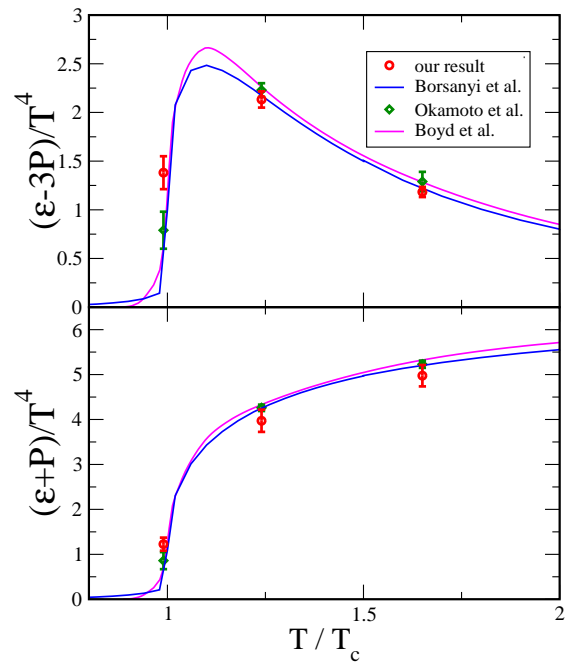}
\caption{Continuum limit of the interaction measure and entropy density
obtained by the gradient flow for~$T/T_c=1.65$, $1.24$, and~$0.99$ with $300$
gauge configurations. The magenta, green, and blue data are the results of the integral method according to~Ref.~\cite{Boyd:1996bx, Okamoto:1999hi, Borsanyi:2012ve}, respectively} 
\label{fig:comparison}
\end{center}
\end{figure}

Finally, we plot, in~Fig.~\ref{fig:comparison}, the continuum limit
of~$\Delta/T^4$ and~$s/T^3$ obtained by the linear fit of the $N_\tau=6$, $8$,
and~$10$ data (the solid lines) in~Fig.~\ref{fig:cont-lim} for $T/T_c=1.65$,
$1.24$, and~$0.99$. For comparison,
the results of~Ref.~\cite{Boyd:1996bx, Okamoto:1999hi, Borsanyi:2012ve} obtained by the integral method
are shown by the magenta, green, and blue data in~Fig.~\ref{fig:comparison}.
The results of the two different approaches are consistent
with each other within the statistical error.

In this Letter, we have proposed and demonstrated 
 a novel way to study thermal $SU(3)$ gauge theory on the lattice.
 The key ingredient is the conserved and
UV-finite energy-momentum tensor~$T_{\mu\nu}^R(x)$ defined 
from the the UV-finite operators ($U_{\mu\nu}(t,x)$ and $E(t,x)$) 
obtained from the Yang--Mills gradient flow
with the matching coefficients ($\alpha_{U,E}(t)$)~\cite{Suzuki:2013gza}.
From the simulations on~$32^3\times(6,8,10)$ lattices with modest statistics
($300$ gauge configurations), we found that the dimensionless interaction
measure and entropy density, $(\varepsilon-3P)/T^4$ and~$(\varepsilon+P)/T^4$,
show plateau structure inside the fiducial window ($2/N_\tau<\sqrt{8t}T<1/2$)
with small statistical errors, so that the double extrapolation~$(a,t)\to(0,0)$ can be
taken appropriately for given~$T$.

Major advantages of the gradient flow applied to the lattice thermodynamics 
 are as follows: (i)~One can simulate $\varepsilon$ and~$P$
independently at any fixed~$T$ through the direct measurement of the
well-defined EMT. There is no need of integration by~$\beta$ or~$T$, which
requires a boundary condition and the numerical interpolation. (ii)~There is no
need of constant subtraction in entropy density~$s$. The interaction
measure~$\Delta$ needs one subtraction of its $T=0$ value, which is obtained by
the accurate measurement of~$t^2\langle E(t,x)\rangle_0$ or by its perturbative
evaluation at small~$t$. (iii)~The statistical noise is substantially reduced
at finite flow time~$t>0$ due to the effective smearing of the operators with
the radius~$\simeq\sqrt{8t}$, so that the extrapolation of the results back
to~$t=0$ is well under control.

Although we studied only the thermal average of EMT in this Letter,
 there is
no conceptual difficulties in applying our method to $n(\ge 2)$-point EMT correlations 
~\cite{Suzuki:2013gza}. This opens the door to investigate transport
coefficients (such as shear and bulk viscosities), fluctuation observables in
the hot plasma, glueballs at zero and finite temperatures. Here we note
 that there is no difficulty in measuring thermodynamic
quantities even at extremely high temperature in this method since no
temperature integration is necessary. It is also an interesting direction to
study the dilation mode or the $a$-function of (nearly) conformal
theory~\cite{Appelquist:2010gy,Latorre:1997ea} using the present method.
Furthermore, including fermions in the present framework extends the scope even
further~\cite{Makino-Suzuki}. Some of these issues as well as the simulations with finer lattice
with larger volume are already started and will be reported elsewhere.

We would like to thank S.~Aoki, F.~Karsch, M.~L\"uscher, and H.~Nagatani for
useful discussions and comments. We are also grateful to H.~Matsufuru for his
help of the code development. Numerical simulation was carried out on NEC SX-8R
and SX-9 at RCNP, Osaka University, and Hitachi SR16000 at KEK under its
Large-Scale Simulation Program (Nos.~T12-04 and 13/14-20). M.~A.,
M.~K., and H.~S. are supported in part by a Grant-in-Aid for Scientific
Researches~23540307, 25800148, and 23540330, respectively. E.~I. is supported
in part by Strategic Programs for Innovative Research (SPIRE) Field~5. T.~H.
 is  supported by RIKEN iTHES Project.

\end{document}